\documentclass[aps,prb,showpacs,tightenlines,twocolumn,secnumarabic,nofootinbib,nobibnotes,groupedaddress]{revtex4-1}

\usepackage{amsmath,amssymb,amsfonts,bm}
\usepackage{graphicx}
\usepackage{dcolumn}
\usepackage[colorlinks=true,linkcolor=blue]{hyperref}%

\begin{document}

%-----------------------------------------------------------------
% documentation    title,  authors,   abstract,   pacs
%-----------------------------------------------------------------

\title{Electrically controllable magnetic order in the bilayer Hubbard model on honeycomb lattice
--- a determinant quantum Monte Carlo study}
%\{}

\author{Jinhua Sun, Donghui Xu, Yi Zhou, and Fu-Chun Zhang}

\affiliation{
Department of Physics, Zhejiang University, Hangzhou, China
}

\date{\today}

\begin{abstract}
Layered antiferromagnetic spin density wave (LAF) state is one of the
plausible ground states of charge neutral Bernal stacked bilayer graphene. In this paper, we use  determinant quantum Monte Carlo method to study the effect of the
electric field on the magnetic order in bilayer Hubbard model on a honeycomb lattice.
Our results qualitatively support the LAF ground state found in the mean field theory.  The obtained magnetic moments, however, are much smaller than what are estimated in the mean field theory. As electric field increases, the magnetic order parameter rapidly decreases.
\end{abstract}

 % set the same as Half-metallic BLG paper

\maketitle

%-----------------------------------------------------------------
% The body of the paper
%-----------------------------------------------------------------

\section{Introduction}

Following the fabrication of monolayer graphene (MLG), \cite{Novoselov04} bilayer graphene (BLG) has attracted intensive attention due to its unique electrical flexibility \cite{Castro07,Oostinga07,Zhang09,Mak09} and unusual physical properties such as unconventional integer quantum Hall effect (where the zero-level Hall plateau is missing).\cite{Novoselov06,McCann06}  If we neglect electron-electron interactions, theoretical model for the bilayer graphene may be  described by a tight-binding Hamiltonian, leading to parabolic valence and conduction bands touching at the highly symmetric Dirac points $K$ and $K^{\prime}$\cite{rmp}, and a band gap may be opened  by
a perpendicular electric field, which saturates at $\sim 0.3eV$, as predicted and also observed in experiments. \cite{Ohta06,McCann06,Min07,Zhang09}

 Most recently, several well controlled experiments\cite{Jr2012,Freitag12,Veligura12,Freitag13,Bao12} have demonstrated that an intrinsic gap of approximately 2meV may exist at the charge neutrality point (CNP) in ultraclean bilayer graphene. Bilayer graphene has finite density of states (DOS) at the CNP,\cite{Castro09} which make it susceptible to the electron-electron interactions.\cite{sun09,Uebelacker11}
Among various broken symmetry ground state candidates,\cite{Nilsson06,Min08,Weitz10,Nandkishore10,Vafek10,Vafek210,Zhang10,Zhang11,Lemonik12,Scherer12,Kotov12,Maxim12,Gorbar12,Zhangfan12} the layered antiferromagnetic (LAF)\cite{Zhang11,Zhang12,Xu12,Scherer12,Wang13,Assaad} spin density wave state is the most probable one. It is interesting to note that an electric field perpendicular to the graphene layers has been applied in experiments to control the physical properties.  So far most theoretical works are based on mean field theories\cite{ZhangJH14,Xu12,Assaad,Yuan13,Murrary13,Yan12,Vafek12,Maxim2012} or renormalization group methods.\cite{Assaad, Murrary13,Vafek10,Vafek12,Honercamp12,Song12,Vafek12,Macdonald12,Lemonik12,Scherer12}  Quantum Monte Carlo technique \cite{Assaad} has been used to study zero temperature properties in the absence of electric fields.

In this paper we use determinant quantum Monte Carlo (DQMC) method to study the effect of the electric field to a bilayer single band Hubbard model\cite{Hubbard63} on a honeycomb lattice, as shown in Fig. \ref{Fig:1_BLG}, relevant to bilayer graphene.

The DQMC method\cite{BSS,Hirsch} we adopt here has been used successfully to study interacting fermion problems.
In this paper, we perform DQMC simulations to study the magnetic order at CNP and the behavior of order parameter when an perpendicular electric field is applied. The DQMC results confirm a LAF ground state at CNP, and the magnetization  monotonically decreases as the potential bias between the two layers  increases.
We also use the self-consistent mean-field analysis to draw the explicit results of critical values of electric potential when the system undergoes a transition from the LAF state to a layered charge polarized (LCP) state, with electron charge imbalance in the two layers and also on the two sublattices in the same layer.

The paper is organized as follows. In Sec. \ref{sec:Model}, we give a brief introduction to our theoretical model and the physical quantities we study. In Sec. \ref{sec:QMC}, we present our results obtained from the DQMC simulations. Firstly, we study the magnetization at the CNP. The purpose is to identify the magnetic order at zero electric field. Then we proceed to discuss the suppression of the magnetization due to the electric field effect. In Sec. \ref{sec:Mean}, we present the self-consistent mean-field results as a comparison and supplementary. Finally in Sec. \ref{sec:Conclusion} we summarize our results.

\section{Model Hamiltonian}\label{sec:Model}
\begin{figure}[htpb]
\begin{center}
\includegraphics[width=8.0cm]{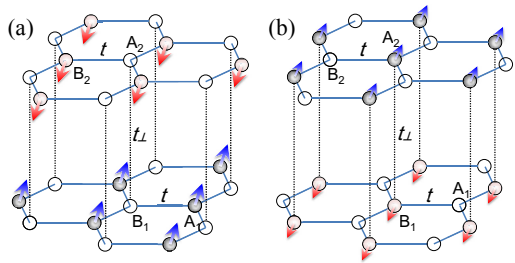}
\end{center}
\caption{(Color online). Bernal stacked bilayer graphene lattice and restricted AFM orders.
A and B denote two sublattices. The subscripts 1 and 2 are layer indices.
Intralayer hopping $t$ links two nearest neighbor sites within the same layer,
while interlayer hopping $t_{\perp}$ links two nearest neighbor sites in opposite layers.
The coordination number $z=3$ for A$_1$ and B$_2$ sites, and $z=4$ for A$_2$ and B$_1$ sites.
(a) Restricted AFM order within $z=3$ sites. (b) Restricted AFM order within $z=4$ sites.
}\label{Fig:1_BLG}
\end{figure}

%We consider a bilayer honeycomb system in an perpendicular electric field. The Hubbard model Hamiltonian is given by
We utilize the Hubbard model to study electron interaction and magnetism in Bernal stacked bilayer graphene. In the presence of an external perpendicular electric field,
the model Hamiltonian contains three parts, the kinetic energy $H_0$, the on-site Coulomb repulsion $H_U$, and the electrostatic potential $H_p$ induced by the
external perpendicular electric field, and reads
\begin{eqnarray}
H=H_0 + H_U + H_p.
\end{eqnarray}
%where $H_0$ is the kinetic energy part, $H_U$ describes the on-site Coulomb repulsion,
%and $H_p$ is the electric potential induced by the external perpendicular electric field.
Here the kinetic energy $H_0$ is given by
\begin{equation}\label{Eq:Hubbard}
\begin{aligned}
H_{0}& =-t\sum_{l\langle ij \rangle\sigma}
[a^{\dag}_{l\sigma}(i)b_{l\sigma}(j) + H.c.] \\
& -t_{\perp}\sum_{\langle i i'\rangle\sigma}[b^{\dag}_{1\sigma}(i)a_{2\sigma}(i') + h.c.] - \mu \sum_{li\sigma}n_{l\sigma}(i),
\end{aligned}
\end{equation}
where $a_{l\sigma}(i)$ ($b_{l\sigma}(i)$) annihilates an electron with spin $\sigma$ at site $i$ in sublattice A (B),
$l=1,2$ is the layer index, $\langle \cdots\rangle$ denotes an intralayer or interlayer nearest neighbor bond as illustrated in Fig. \ref{Fig:1_BLG}.
To be simple, we only take into account the nearest neighbor intralayer hopping $t$ and the nearest neighbor interlayer hopping $t_\perp$.
$n_{l\sigma}(i)=a_{l\sigma}^{\dagger}(i)a_{l\sigma}(i)$ or $(b_{l\sigma}^{\dagger}(i)b_{l\sigma}(i))$, depending on which sublattice (A or B) the site $i$
belongs to. $\mu$ is the chemical potential, and $\mu=0$ corresponds to the charge neutrality point.
The on-site Hubbard interaction $H_U$ can be written as
\begin{equation}
\begin{aligned}
H_U &=U\sum_{li}[n_{l\uparrow}(i)-1/2][n_{l\downarrow}(i)-1/2],
\end{aligned}
\end{equation}
where $U>0$ for the repulsion. The effect of the applied perpendicular electric field is parameterized by the potential difference $V$ between the two layers,
\begin{equation}
\begin{aligned}
H_p &=\sum_{li \sigma} V_l n_{l\sigma}(i),\\
\end{aligned}
\end{equation}
where $V_l=(-1)^lV/2$.

The intralayer nearest neighbor hopping $t$ is estimated about $3.16eV$ in bilayer graphene.
For convenience, we shall use $t$ as the energy unit in our calculation.
The interlayer nearest neighbor hopping $t_{\perp}$ is commonly used as $0.381eV\sim0.13t$ in literature.
However, the effective parameters of bilayer graphene used in model studies are not given $a$ $priori$,
which is one of the roadblocks for determining the exact ground state of bilayer graphene.
In this context, we would rather consider about more general situation: the parameters chosen are unrealistic,
but more interesting than the real bilayer graphene parameter regime.
In this paper, we shall consider two values for the interlayer hopping energy: $t_{\perp}=0.2t$ and $1.0t$.

We close this section by providing an alternative site labeling scheme for the later use.
Note that each unit cell in the Bernal stacked bilayer graphene contains four sites (two layers and two sublattices).
The Bravais lattice for such a system is indeed a triangular lattice. Then one can denote a site as $(\vec{R},d)$, where $\vec{R}$ labels a unit cell
and $d$ labels the atom site within a unit cell, say, basis site.
As would be seen in the next section, this Bravais lattice + basis parametrization is more convenience to study the magnetic order and correlation.

\section{Quantum Monte Carlo Simulations}\label{sec:QMC}

%We apply the DQMC method to obtain unbiased numerically exact results of the magnetic order.
We apply the DQMC method to study spin correlations and magnetic susceptibility in bilayer graphene, which avoids any assumption on the magnetic structure as
in mean field theory or other approximations. In this sense, it is unbiased and numerically exact, namely, it is free of systematic errors but not random errors.
DQMC evaluates the imaginary time Green's functions and thereby various correlation function through Wick's theorem at finite temperature.

As a finite temperature method, DQMC is able to examine different types of instability as temperature lowering.
At first glance, this method is not applicable to catch possible long-ranged magnetic orders in the ground state.
Since the Mermin-Wagner theorem\cite{Mermin} does not allow any long-ranged orders in the two dimensional Hubbard model at finite temperature.
However, as pointed by Hirsch,\cite{Hirsch} if we go to sufficiently low temperature where the thermal coherence length, or the thermal de Broglie wavelength is much larger than
the linear system size, the system behaves as if it is at zero temperature. This makes DQMC able to catch ground state features in two dimension.

%The DQMC simulations naturally return the imaginary time Green's functions. Using these Green's functions and basing on the Wick's theorem,
%we can obtain various correlation functions. Our method can only deal with finite temperature, and the 2-dimensional Hubbard model cannot exhibit long-range magnetic order
%according to the Mermin-Wagner theorem\cite{Mermin}. However, if we go to sufficiently low temperature that the thermal coherence length is much larger than the linear system size,
%our procedure will work since the system behaves as if at zero temperature.\cite{Hirsch}

To study spin correlations and magnetic susceptibility in bilayer graphene, we choose the four-atom unit cell as discussed in the last paragraph in the last section.
Thus both the spin structure factor and the spin susceptibility can be described by matrix elements of $4\times 4$ matrices.
The $zz$ component of the static spin structure factor ($\omega=0$) is defined as
        \begin{eqnarray}\label{Eq:SAF}
        S_{dd'}(\vec{q})\equiv \frac{1}{L^2}\sum_{\vec{R},\vec{R}^{\prime}}  e^{i\vec{q}\cdot (\vec{R}-\vec{R}^{\prime})} \langle S_{\vec{R}d}^{z}\cdot S_{\vec{R}^{\prime}d'}^{z}\rangle,
        \end{eqnarray}
where $L$ is the linear size of the Bravais lattice, thereby $L^2$ is the number of unit cells, and the total site number in the bilayer system is $N=4L^2$.
The indices $d$ and $d'$ denote the basis sites within a unit cell $\vec{R}$ or $\vec{R}^{\prime}$.

By diagonalizing the $4 \times 4$ matrix $S_{dd'}(\vec{q})$, we obtain the maximum eigenvalue $S(\vec{q})$. This spin structure factor $S(\vec{q})$
describe the dominant bilayer magnetic correlation at wave vector $\vec{q}$. We shall call it ``dominant spin structure factor''.
%At sufficiently low temperature and large system size, the staggered magnetization per lattice site is given by
Correspondingly, we can define a dominant magnetization (or staggered magnetization, depending on the ground state) per site as follows,
$$m=\sqrt{S(q=0)/4L^2}.$$

Beside the dominant spin structure factor $S(\vec{q})$, we can also define the restricted structure factors for sites with coordination numbers $z=3$ ($A_1$ and $B_2$)
and $z=4$ ($B_1$ and $A_2$),
        \begin{eqnarray}\label{Eq:SAFz}
        S_{z}\equiv \frac{1}{2L^2}\sum_{i,j|z_i=z_j=z} \epsilon_{i,j} \langle S_{i}^{z}\cdot S_{j}^{z}\rangle,
        \end{eqnarray}
from which we can obtain the local order parameters,
$$m_z=\sqrt{S_{z}/(2L^2)}.$$
If the two sites $i$ and $j$ belong to the same sublattice (A or B), then
$\epsilon_{i,j}=1$, otherwise $\epsilon_{i,j}=-1$. The corresponding magnetic orders are shown in Fig. \ref{Fig:1_BLG}(a) and Fig. \ref{Fig:1_BLG}(b) respectively.

Similarly, the static magnetic susceptibility is given by a $4\times4$ matrix,
        \begin{eqnarray}
        \chi_{dd'}(\vec{q})\equiv \frac{1}{L^2}\sum_{\vec{R},\vec{R}^{\prime}} e^{i\vec{q}\cdot (\vec{R}-\vec{R}^{\prime})}\int_{0}^{\beta}\langle S_{\vec{R}d}^{z}(\tau)\cdot S_{\vec{R}^{\prime}d'}^{z}\rangle.
        \end{eqnarray}
The behavior of the largest eigenvalue of the matrix $\chi_{dd'}(\vec{q})$, $\chi(\vec{q})$, reveals the dominant magnetic response of the bilayer system.

As mentioned in previous section, we mainly consider two values of $t_\perp$ in our calculations,
(1) $t_\perp=0.2t$ which is close to the realistic value of bilayer graphene and
(2) $t_\perp=1.0t$ where the symmetry breaking among the sublattices plays a more significant role.
Using these parameters, we carry out DQMC calculation on $L\times L\times 4$ lattices up to $L=9$ to study magnetic orders and fluctuations
as the on-site Coulomb repulsion $U$ and the interlayer bias $V$ vary.

Again, we would like to emphasize that DQMC is a finite temperature method, according to Hirsch's argument\cite{Hirsch},
it can catch ground state features on finite size lattices only when the thermal coherence length is larger than the linear system size.
For this purpose, we scale the temperature with the system size,
and the temperature is chosen as $1/T=1.2L$. $L=9$ is the largest linear system size used in our DQMC simulations.
The corresponding temperature is set as $T=1/10.8$, which is about the lowest temperature we can approach without numerical instability.

%We first present the numerical results of $S(q=0)/4L^2$ while potential bias $V$ is set to be zero.
%The purpose is to identify the magnetic order at zero electric field,
%and only after that can we proceed to study the effect of external electric field on the magnetic property of the bilayer system.

\subsection{Magnetic ordering in the absence of external electric field}

\begin{figure}[htpb]
\begin{center}
\includegraphics[width=8.0cm]{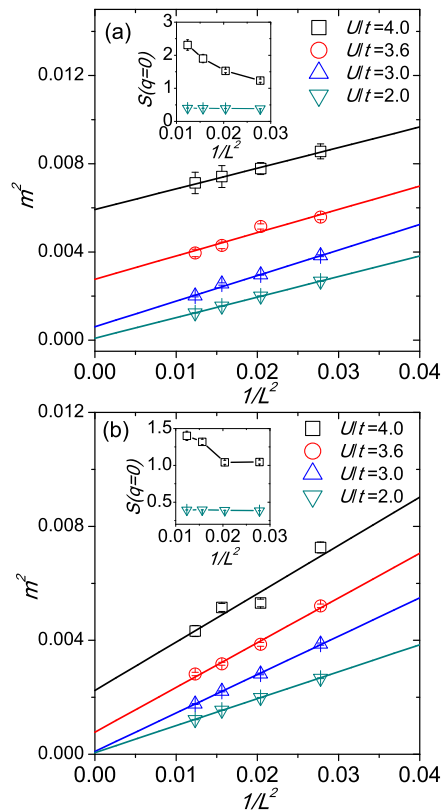}
\end{center}
\caption{(Color online). The dominant magnetization square $m^2$ is plotted as a function of $1/L^2$ for different values of $U$.
Here we set $1/T=1.2L$ and $V=0$. (a) $t_\perp=1.0t$, (b) $t_\perp=0.2t$.
The symbols are numerical results of DQMC simulations, and the lines are linear fitting.
Insets: $S(q=0)=4m^2L^2$ is plotted as a function of $1/L^2$, from which it is clear seen that $m(L\to\infty)=0$ at $U/t=2.0$.
  }\label{Fig:S_diffT}
\end{figure}

Firstly, we would like to examine numerically the existence of magnetic order in the absence of external electric field.
Only after that can we proceed to study the electric control of magnetic order in the bilayer graphene.
In order to do this, we calculate the dominant magnetization $m$ and extrapolate it to the thermodynamic limit $L\to\infty$.
A finite extrapolated value of $m$ at thermodynamic limit indicates an magnetic ordered ground state.
We present numerical results for $t_\perp=1.0t$ and $t_\perp=0.2t$ in Fig.\ref{Fig:S_diffT}.

%In Fig. \ref{Fig:S_diffT}, we present the results of $S(q=0)/4L^2$ versus $1/L^2$ for different values of $U$.
%By extrapolating the results to $N\rightarrow \infty$, we can obtain the magnetization $m=\sqrt{S_{q=0}/4L^2}$, and a finite value of $m$ at large $N$ limit
%indicates an ordered ground state.

As shown in Fig. \ref{Fig:S_diffT} (a), when the interlayer hopping $t_{\perp}=1.0t$, the dominant magnetization $m$ vanishes at $U/t=2.0$ but becomes finite
at $U/t=3.0,3.6,4.0$ in the thermodynamic limit. The critical point $U=U_c$ separating the magnetic ordered phase and the paramagnetic phase is in the range of
$2.0t<U_c<3.0t$. Fig. \ref{Fig:S_diffT} (a) shows the results for weaker interlayer hopping $t_{\perp}=0.2t$, which is close to the realistic value
in bilayer graphene. It is clear that $3.0t<U_c<3.6t$ when $t_{\perp}=0.2t$. Comparing the estimated ranges for $U_c$ at $t_{\perp}=1.0t$ and $t_{\perp}=0.2t$
as well as the absolute value of $m$, one finds that the interlayer hopping $t\perp$ favor magnetic ordering states and will enhance magnetic correlations.
Since DQMC is a finite temperature method, thermal fluctuations will suppress the long ranged magnetic order, it must lowerestimate the value of $m$
for ground state. But this deviation is not significantly when the temperature is much smaller than the energy gap. For $U=4.0t$, the energy gap is about $0.3t$,
the dominant magnetization we obtain by DQMC simulations at $T=1/10.8$ is $\sim 0.12$, which is in good agreement with that obtained by projector QMC at zero temperature.\cite{Assaad}

\begin{figure}[htpb]
\begin{center}
\includegraphics[width=8.0cm]{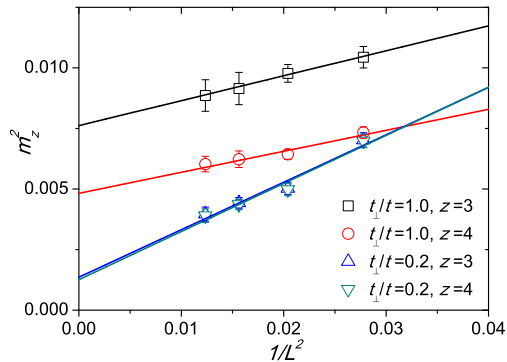}
\end{center}
\caption{(Color online). The restricted magnetization square $m_{z}^2$ in the absence of external electric field is plotted as a function of $1/L^2$.
Here we set $U=4.0t$. }\label{Fig:S_diff_tper}
\end{figure}

Moreover, we can specify the magnetic structure by studying the local order parameters $m_z$. As shown in Figs. \ref{Fig:1_BLG}(a) and \ref{Fig:1_BLG}(b),
$m_{z=3}$ is the staggered magnetization in the $z=3$ sublattices (A$_1$ and B$_2$),
while $m_{z=4}$ is the staggered magnetization in the $z=4$ sublattices (A$_2$ and B$_2$).
The later two sublattices ($z=4$) are linked by the interlayer hopping $t_{\perp}$.
As plotted in Fig. \ref{Fig:S_diff_tper}, for $V=0$ and $U=4.0t$, both staggered magnetization $m_{z=3}$ and $m_{z=4}$ exist when $t_{\perp}=0.2t$
or $t_{\perp}=1.0t$. For $t_{\perp}=0.2t$, the values of $m_{z=3}$ and $m_{z=4}$
almost coincide each other, while for $t_{\perp}=1.0t$, $m_{z=4}$ is apparently smaller than $m_{z=3}$ although both $m_{z=4}$ and $m_{z=3}$ are larger than their
values at $t_{\perp}=0.2t$.

The similarity and difference between $t_{\perp}=0.2t$ and $t_{\perp}=1.0t$ can be explained as follows.
Both $m_{z=3}$ and $m_{z=4}$ reflect antiferromagnetic spin correlation between the two layers.
Increasing $t_{\perp}$ will enhance this correlation directly or indirectly, thus both $m_{z=3}$ and $m_{z=4}$ will increase with $t_{\perp}$
as seen in Fig. \ref{Fig:S_diff_tper}.
However, the other tendency induced by the hopping terms (both $t$ and $t_{\perp}$) is to against the on-site Coulomb repulsion $U$ and drive the electrons to
be itinerant rather than localized. This will reduce the magnitude of local magnetic moment effectively. Therefore the magnitude of the local moment at $z=4$ sites becomes
smaller than that at $z=3$ sites, resulting in $m_{z=4}<m_{z=3}$.

%Fig. \ref{Fig:S_diff_tper} shows the results of $S_{z}/2L^2$ versus $1/L^2$ for $t_\perp=1.0t$ and $0.2t$. The values of parameters are chosen as
%$U=4.0t$, and the potential bias is still zero. At $t_\perp=0.2$, the interlayer hopping is small compared to the intralayer hopping $t$.
%Hence the interlayer hopping causes symmetry breaking between the two sublattices in the same layer, but the effect is not significant.
%It can be learned from our numerical results that the values of two restricted structure factors for $t_\perp=0.2t$ nearly coincide.
%However, as $t_\perp$ increases to $1.0$, the distinction between the two restricted structure factors becomes significant.
% The magnetization on the sites with $z=3$ as well as $z=4$ are increased.
%For both $t_\perp=0.2t$ and $1.0t$, the results shown in Fig. \ref{Fig:S_diffT} and Fig. \ref{Fig:S_diff_tper}
%show that long-range LAF state is formed in the bilayer honeycomb system while $U$ is set to be $4.0t$.

\subsection{Magnetic orders and fluctuations in the presence of external perpendicular electric field}

\begin{figure}[htpb]
\begin{center}
\includegraphics[width=8.0cm]{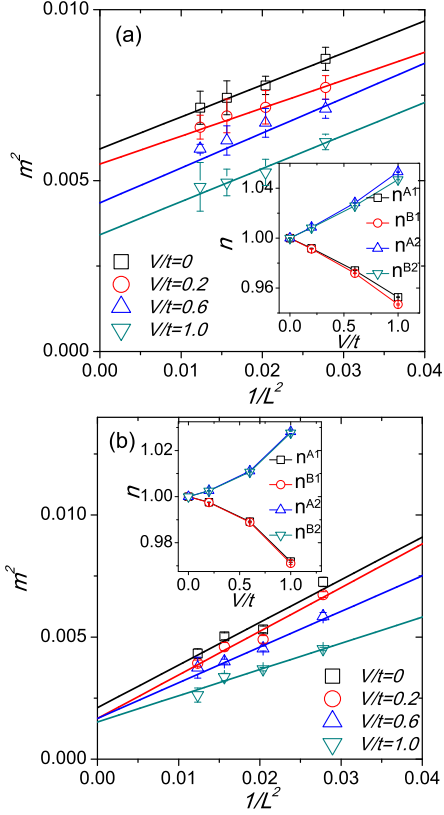}
\end{center}
\caption{(Color online). The dominant magnetization square $m^2$ is plotted as a function of $1/L^2$ in the presence of external perpendicular electric field.
Here we set $U=4.0t$. (a) $t_\perp=1.0$, (b) $t_\perp=0.2$.
Insets: The electron occupation numbers in four sublattices.}\label{Fig:S_diffV}
\end{figure}

Now we turn on the external perpendicular electric field to see its effect on the LAF states.
In bilayer graphene, experimentally tunable bias $V$ is up to about $1.0t$. In principle, we can use larger $V$
in theoretical study. However, our DQMC simulation will suffer from serious negative sign problem when $V>1.0t$, see Appendix for details.
We find that the dominant magnetization $m$ will be suppressed by the external perpendicular electric field.
The numerical results are summarized in Fig. \ref{Fig:S_diffV}. We fix $U=4.0t$ and focus on two values of $t_{\perp}$, $1.0t$ and $0.2t$,
where LAF order is well established in the absence of external electric field. Fig. \ref{Fig:S_diffV} exhibits
clear tendency that long ranged magnetic order will be suppressed by the perpendicular electric field. However, $V=1.0$eV, which is the maximum
value approachable by our DQMC simulation, is still insufficient to destroy the magnetic ordering entirely.
We shall apply the mean field theory to study the possible transition from the LAF ordered phase to the paramagnetic phase
in the next section for completeness.

%While $t_\perp=0.2$, Fig. \ref{Fig:S_diffV} (b) shows monotonically decreasing magnetization as that in Fig. \ref{Fig:S_diffV} (a).
%Although the magnetization for $t_\perp=0.2$ is much smaller than that in $t_\perp=1.0$ at zero external field, $V=1.0$ is still not
%large enough to completely destroy the magnetization, so we will apply the self-consistent mean-field analysis to study the critical values of $V$
%when the system undergoes a transition from a paramagnetic to an LAF states.

\begin{figure}[htpb]
\centering
\includegraphics[width=8.0cm]{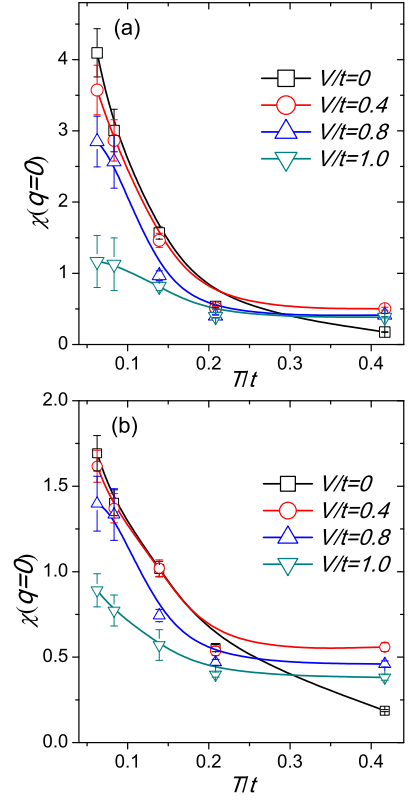}
\caption{(Color online). The spin susceptibility $\chi$ as a function of $T$ in the presence of external perpendicular electric field.
Here we set $U=4.0t$. (a) $t_\perp=1.0$, (b) $t_\perp=0.2$.  }\label{Fig:X_diff_V}
\end{figure}

Magnetic fluctuations are investigated through the temperature dependent static spin susceptibility $\chi(\vec{q}=0)$.
We still set $U=4.0t$ and study the cases of $t_\perp=1.0t$ and $t_\perp=0.2t$.
It is shown in Fig. \ref{Fig:X_diff_V} that the magnetic susceptibility $\chi$ will diverge at $V=0$ as temperature $T\to 0$.
This instability will give rise to magnetic ordering at zero temperature.
As the potential bias $V$ increases, the magnetic susceptibility $\chi$ will be gradually suppressed.

From Fig. \ref{Fig:S_diffV} and Fig. \ref{Fig:X_diff_V}, one sees that both magnetic ordering and magnetic fluctuations are suppressed by the
applied perpendicular electric field. This means that the magnitude of local magnetic moment is reduced. The reduction of local moment can be
explained as follows. At half filling, the average electron occupation number per site is unit. The perpendicular electric field will increase
the electron occupation number on one layer and decrease that on the other layer, resulting in reduced local moment.
This is in agreement with the deviation of electron occupation number from the unit as shown in the insets in Fig. \ref{Fig:S_diffV}.
We shall also examine this argument through mean field theory in next section.

%Fig. \ref{Fig:X_diff_V} shows the magnetic susceptibility $\chi$ as a function of $T$ for various values of potential bias $V$, we use $U=4.0$.
%The figures from top to bottom is arranged in the order of increasing potential bias $V=0, 0.4t, 0.8t, 1.0t$.
%Shown in Fig. \ref{Fig:X_diff_V} (a) are the results for $t_\perp=1.0$. We can see that when there is no potential bias added to the bilayer system, the magnetic susceptibility $\chi$ tends to diverge
%as the temperature approaches zero, which indicates the formation of long-range magnetic order in this system.
%However, as the potential bias increases, the magnetic susceptibility is much suppressed,
%and as $T\rightarrow 0$, the susceptibility tends to be finite.
%This means that the long-range magnetic order is suppressed due to the electric potential.
%The results for $t_\perp=0.2$ are shown in Fig. \ref{Fig:X_diff_V} (b). For $V=0$, the spin susceptibility also diverge at low temperature, but
%with much smaller values compared with those for $t_\perp=1.0$. As the potential bias increases to $V=1.0$,
%the spin susceptibility is gradually reduced.
%The non-divergent behavior of spin susceptibility at low temperature as $V$ increases indicates that the magnetic order is
%destroyed by the electric field applied in the bilayer system.

\section{Mean Field Theory}\label{sec:Mean}

In order to go beyond the limit of DQMC, where the reliable results can be obtain only when $V<1.0t$ due to sign problem, we carry out mean field analysis
to investigate possible phase transition from the SDW state to paramagnetic state as $V$ increases.
Following Ref. \cite{Xu12}, we introduce the following mean field decomposition for the Coulomb interaction term,
\begin{equation}
H_{U}^{MF}=U\sum_{l,i\sigma}[(\langle n_{l\sigma}(i)\rangle-1/2)(n_{l\bar\sigma}(i)-1/2)],
\end{equation}
where $\bar\sigma=-\sigma$, and $\langle n_{l\sigma}(i)\rangle$ should be determined self-consistently.
To describe LAF spin density wave states, we introduce eight mean fields $\langle n_{l\sigma}^{\eta}\rangle$ for $\langle n_{l\sigma}(i)\rangle$,
where $\eta$ indicates the sublattice A or B. Then local magnetization and electron occupation number at each carbon atom site can be defined
in terms of $\langle n_{l\sigma}^{\eta}\rangle$ respectively,
\begin{equation}
m_{l}^{\eta}=\frac{1}{2}\langle n_{l\uparrow}^{\eta}-n_{l\downarrow}^{\eta}\rangle
\end{equation}
and
\begin{equation}
n_{l}^{\eta}=\langle n_{l\uparrow}^{\eta}+n_{l\downarrow}^{\eta}\rangle.
\end{equation}

%In this section, we present the mean-field results of magnetization for various values of potential bias $V$.
%Detailed mean-field analysis applied on a bilayer graphene Hubbard model with electric field can be found in Ref. \cite{Xu12}, the results we show in this context
%are for comparison and supplementary since the DQMC simulations cannot give the exact values of critical potential bias.

%We use a mean-field approximation for the Hubbard term represented in Eq. \ref{Eq:Hubbard}, and solve the Hamiltonian self-consistently
%\begin{equation}
%\begin{aligned}
%H_{U}^{MF}=U\sum_{li\sigma}[(\langle n_{l\sigma}\rangle-1/2)(n_{l\sigma^{\prime}}-1/2)].
%\end{aligned}
%\end{equation}
%In the honeycomb bilayer system, each unit cell consists of four atoms, which means that we have to deal with eight mean-fields,
%$\langle n_{l\sigma}^{\eta}\rangle$, with $\eta$ indicating the sublattice A or B.

\begin{figure}[htpb]
\begin{center}
\includegraphics[width=8.0cm]{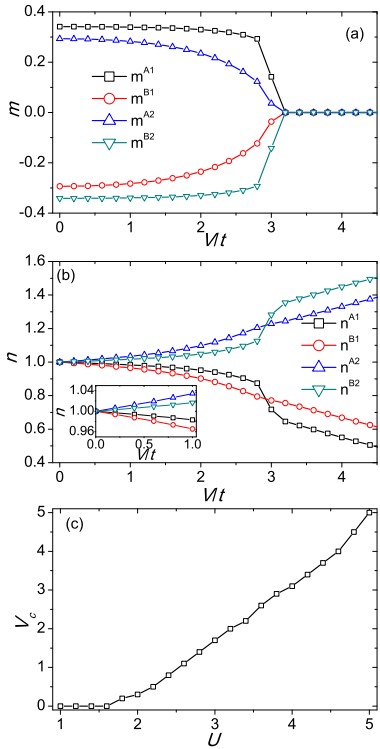}
\end{center}
\caption{(Color online). (a) The mean field results of magnetization $m$ and (b) charge number for $U=4.0t$ and $t_\perp=1.0t$ as a function of potential bias $V$. Here we use $U=4.0$.
Inset: A zoom for the charge number at the range of $0<V<1.0t$.
(c) The critical potential bias $V_c$ for various values of $U$ for $t_\perp=1.0t$.
 }\label{Fig:meanfield}
\end{figure}
%\subsection{Effect of $t_{\perp}$}

We choose $t_{\perp}=1.0t$ and $U=4.0t$ to study how the magnetization will be reduced and
how charge imbalance between the two layers will be induced as the potential bias $V$ increasing.
The mean field results are shown in Fig. \ref{Fig:meanfield}(a) and \ref{Fig:meanfield}(b).

When $V=0$, the ground state is a gapful LAF state. As shown in Fig. \ref{Fig:meanfield}(a),
the magnitude of the magnetization at $z=3$ sites (A$_1$ and B$_2$) is larger than that at $z=4$ sites (A$_2$ and B$_1$),
which is consistent with DQMC results (see Fig. \ref{Fig:S_diff_tper}). However, the magnitude of the magnetization
given by mean field theory is bigger than those in DQMC simulation. This is because of two reasons.
Firstly, DQMC is a finite temperature method, thermal fluctuations are involved, which will reduce order parameters
on the ground state. Secondly, the mean field theory neglects the quantum fluctuations and tends to overestimate the magnetization.

As the potential bias $V$ increasing, the system experiences continuous charge transfer from one layer to the other as shown in \ref{Fig:meanfield}(b).
This charge transfer will induce charge imbalance between the two layers.
Since $t_{\perp}$ only links sublattices A$_2$ and B$_1$, it will also induce charge imbalance between sublattices A and B.
The charge transfer will also reduce the local magnetic moment effectively.
The larger charge imbalance (measured from single occupancy) corresponds to smaller local moment.
In the large $V$ limit, the system becomes a paramagnetic and layered charge polarized state. The phase transition from LAF to paramagnetic state
happens at $V=V_c\approx 3.2t$.

It is interesting that although the charge imbalance between two layers and two sublattices drives each site away from single occupancy,
the relation
\begin{equation}\label{Eq:nAB}
n^{A_1}+n^{B_2}=n^{B_1}+n^{A_2}=2.0
\end{equation}
still keeps up. This relation can be explained by a combined symmetry. Although
the external electric field breaks the charge conjugation (or particle-hole) symmetry as well as the lattice inversion symmetry,
the combination of the charge conjugation and the lattice inversion is still a symmetry operation under the external electric field.
This combined charge-conjugation-lattice-inversion symmetry will guarantee Eq. (\ref{Eq:nAB}).

Finally, we fix the interlayer hopping $t_{\perp}=1.0t$ and vary on-site Coulomb repulsion $U$
to study the critical values of $V$, which separates the LAF phase and the paramagnetic phase.
The mean field results are present in Fig. \ref{Fig:meanfield} (c).
By the mean field theory, the system starts to develop the magnetization at $U\sim 1.8t$. Thus the value $V_c$ makes sense only when $U> 1.8t$.
In Fig. \ref{Fig:meanfield} (c), one sees that $V_c$ increases monotonically with $U$, and the increasing becomes faster when $U>4.0t$.
$V_c$ is mainly determined by the spin density wave (SDW) gap opened by $U$. Since the SDW gap increases exponentially with $U$ near $U_c$,\cite{Assaad},
we expect that $V_c$ will also grows nearly exponentially with $U$.

%In Fig. \ref{Fig:meanfield} (c), we show the critical potential bias for various values of on-site Coulomb repulsion $U$ while $t_\perp=1.0$.
%The system starts to develop magnetization at $U~1.8t$,
%In Fig. \ref{Fig:meanfield} (c), we can see that $V_c$ increases monotonically with $U$, and $V_c$ increases faster when $U>4.0t$.
%We note that the value of $V_c$ is mainly affected by the single particle gap opened by $U$, and since the gap increases exponentially with $U$,\cite{Assaad},
%we expect that $V_c$ will also grows nearly exponentially with $U$.

\section{Discussion and conclusion}\label{sec:Conclusion}
In conclusion, we have studied the effect of external potential bias on the magnetization in the bilayer honeycomb lattice Hubbard model.
Two typical values of interlayer hopping energy are tested: $t_\perp=0.2t$, which is close to the value in bilayer graphene with Bernal stacking; and $t_\perp=1.0t$,
which represents a larger interlayer coupling.
 In the absence of a potential bias in the bilayer system, there is a layered antiferromagnetic order at $U>3.0t$.
The magnetization on the sites connected by the interlayer hopping "bonds" is suppressed by the interlayer hopping, while the magnetization on other sites is slightly enhanced by the interlayer hopping. For both $t_\perp=1.0t$ and $0.2t$, both of the DQMC results and the mean-field analysis support the LAF ground state.
 In the presence of a perpendicular electric field,  the  antiferromagnetic order is suppressed.
Due to the negative sign problem, we are not able to approach large values of potential bias and to obtain the critical value $V_c$ for the transition from
 an antiferromagnetic to paramagnetic states in the DQMC simulation. However, the tendency of decreasing magnetization as $V$ increases is clearly observed in the QMC results,
and the critical value $V_c$ for various values of $U$ is explicitly given by the self-consistent mean-field analysis.

\section{Acknowledgement}
This work is supported in part by National Basic Research Program of China
(No.2011CBA00103/2014CB921201/2014CB921203), NSFC (No.11374256/11274269), and the
Fundamental Research Funds for the Central Universities in China.


%\appendix
%\section{Derivation of spin-spin interaction}

%-----------------------------------------------------------------
% Sec**: References
%-----------------------------------------------------------------
%\nocite{*}
\appendix
\section{On negative sign problem}

In a Monte Carlo simulation, we evaluate the expectation value of a quantity $A$ as follows,
\begin{equation}\label{Eq:A1}
\langle A\rangle_p = \frac{\sum_{i} p_{i} A_{i} }{\sum_{i} p_{i}},
\end{equation}
where $\sum_{i} p_{i}>0$. If $p_{i}>0$, it serves a ``Boltzmann weight'' and describes a distribution.
Then one can use the distribution $p$ to evaluate the expectation value $\langle A\rangle$ by Monte Carlo method.
However, sometimes $p_{i}$ is not positive definite, namely, $p_{i}<0$ occurs. In this case, $p_{i}$ is no longer a distribution function.
But one can still rewrite $\langle A\rangle_p$ in Eq. (\ref{Eq:A1}) as
\begin{equation}\label{Eq:A2}
\langle A\rangle_p = \frac{\sum_{i} p_{i} A_{i} }{\sum_{i} |p_{i}|}\frac{\sum_{i} |p_{i}| }{\sum_{i} p_{i}}=\frac{\langle sA\rangle_{p'}}{\langle s\rangle_{p'}},
\end{equation}
where $p'=|p|$ and $s_{i}=p_{i}/|p_{i}|$ is the sign of $p_{i}$. The one can evaluate two expectation values $\langle sA\rangle_{p'}$ and $\langle s\rangle_{p'}$
by Monte Carlo method under the distribution $p'$.

If $\langle s\rangle_{p'}$ is sizable and much larger than the statistical error bar, the Monte Carlo simulation is still efficient. However,
when $\langle s\rangle_{p'}\ll 1$, the strong statistical fluctuation cannot be compensated by longer Monte Carlo runs.
This is so called ``negative sign problem''. Actually, it has been confirmed that the average sign has the relation
$\langle s\rangle\sim e^{-\beta N \gamma}$, where $\gamma$ depends on the filling $n$ and interaction $U$.
In our calculation, as the system size grows, and as the potential bias increases, the negative sign problem
becomes quite serious. For example, while $t_\perp=1.0t$, $U=4.0t$ and $T=1/10.8$, the sign average $\langle s\rangle\sim0.1$ for $V=1.0t$
when the system size is $N=9\times 9\times 4$.

\bibliographystyle{apsrev4-1}
\bibliography{ref}

\end{document}